\documentclass[conference,compsoc]{IEEEtran}

\usepackage{cite}
\usepackage{graphicx}
\usepackage{xcolor}
\usepackage{algorithm}
\usepackage[noend]{algpseudocode}
\usepackage{listings}

\renewcommand\tablename{Example}

\makeatletter
\AtBeginDocument{%
	\let\c@table\c@lstlisting
	\let\thetable\thelstlisting
	\let\ftype@lstlisting\ftype@table % give the floats the same precedence
}
\makeatother

\begin{document}
	
\title{Lockless Transaction Isolation in Hyperledger Fabric}

\author{\IEEEauthorblockN{Hagar Meir, Artem Barger, Yacov Manevich, Yoav Tock}
	\IEEEauthorblockA{IBM Haifa Research Lab \\
		Haifa, Israel \\
		hagar.meir@ibm.com, \{bartem, yacovm, tock\}@il.ibm.com}
	}

\maketitle

\begin{abstract}	
%abstract
Hyperledger Fabric is a distributed operating system for permissioned blockchains hosted by the Linux Foundation. It is the first truly extensible blockchain system for running distributed applications at enterprise grade scale. To achieve this, Hyperledger Fabric introduces a novel execute-order-validate blockchain architecture, allowing parallelization of transaction execution and validation. However, this raises the need for transaction isolation. Today transaction isolation is attained by locking the entire state database during simulation of transactions and database updates. This lock is one of the major performance bottlenecks as observed by previous work.

This work presents a new lock-free approach for providing transaction isolation. It harnesses the already existing versioning of key-value pairs in the database, used primarily for a read-write conflict detection during the validation phase, to create a version-based snapshot isolation. We further implement and evaluate our new approach. We show that our solution outperforms the current implementation by 8.1x and that it is comparable to the optimal solution where no isolation mechanism is applied. 

\end{abstract}


\section{Introduction}
%introduction
A \emph{blockchain} is an emerging technology receiving colossal attention over the past decade mainly due to its inherent capability of maintaining a tamper-proof shared distributed ledger among mutually distrusting parties.

Blockchain, also referred to as distributed ledger technology (DLT), merits its popularity from being extremely useful in a wide range of distributed applications due to its unique properties, such as non-repudiation, transparency, provenance, and fault-tolerance. More specifically, blockchain is a distributed, immutable, append-only log of ordered transactions, where order is obtained through a distributed consensus algorithm.
Each transaction is cryptographically signed and transactions are grouped into blocks, to optimize bandwidth utilization~\cite{friedman1997packing}. Blocks are then hash-chained together and recorded as a \emph{ledger}. Each party is responsible of maintaining its own copy of the distributed ledger, assuming that everyone else is not trustworthy. Therefore, any attempt to forge or replace parts of those transactions could be detected, which provides guarantees of data finality and integrity.

% permissionless
Initially blockchain was devised for trusted exchange of digital goods, where one of the first and most prominent examples is Bitcoin~\cite{bitcoin} -- the distributed crypto-currency blockchain system.
Such blockchain systems are known as public or \emph{permissionless} blockchains. In a permissonless blockchain, anyone can join or leave the network, and no one is required to specify its real identity. In such settings no participant can be really trusted. This lack of identification necessitates employing a computationally heavy consensus mechanism which is based on cryptography -- \emph{proof-of-work}.
The proof-of-work consensus protocol has several salient disadvantages: (1) a huge computational cost, resulting in excessive power consumption, (2) probabilistic nature of transaction confirmation, leading to a considerably long confirmation latency, and (3) low transaction throughput. These factors make public blockchains unsuitable for enterprise grade applications.
%

% permissioned
Blockchain further attracts significant attention from enterprises in use cases such as supply chain management, insurance, healthcare, and many others, where business processes run among identifiable participants that are otherwise mutually distrusting~\cite{blockchainInWild}. Therefore, \emph{permissioned} blockchains have naturally emerged as a sane alternative for permissionless networks, being able to address the business needs of enterprise use cases, both in terms of performance and security. In the permissioned setting, a blockchain could be viewed as a traditional replicated state machine (RSM)~\cite{SMR}, where the most natural way to implement RSM is to have a consensus algorithm to decide on order of transactions and then execute them sequentially on each computational node~\cite{lamport1998part}. This is known as the \emph{order-execute} architecture which leads to intolerance of non-determinism in \emph{smart contracts} and to sequential execution of transactions which severely limits performance~\cite{FabricPaper}.

% execute-order-validate
Hyperledger Fabric~\cite{FabricPaper} is an open source project, released by the Linux Foundation\footnote{www.linuxfoundation.org}. It introduces a new architecture for enterprise grade permissioned blockchain platforms following the novel paradigm of \emph{execute-order-validate} for distributed execution of smart contracts (\emph{chaincode} in Hyperledger Fabric). In contrast to the order-execute paradigm, in Hyperledger Fabric transactions are first tentatively \emph{executed}, or endorsed, by a subset of peers. Transactions with tentative results are then grouped into blocks and \emph{ordered}. Finally, a \emph{validation} phase makes sure that transactions were properly endorsed and are not in conflict with other transactions. Valid transactions are then committed to the blockchain state, while invalid transactions are omitted from the state. Note that both valid and invalid transactions are stored in the ledger, and a replay of the validation phase on the ledger deterministically reconstructs the state.
This architecture allows multiple transactions to be executed in parallel by disjoint subsets of peers, increasing throughput, and tolerates non-deterministic chaincode.

\begin{figure*}[ht]
	\centering
	\includegraphics[width=0.7\linewidth]{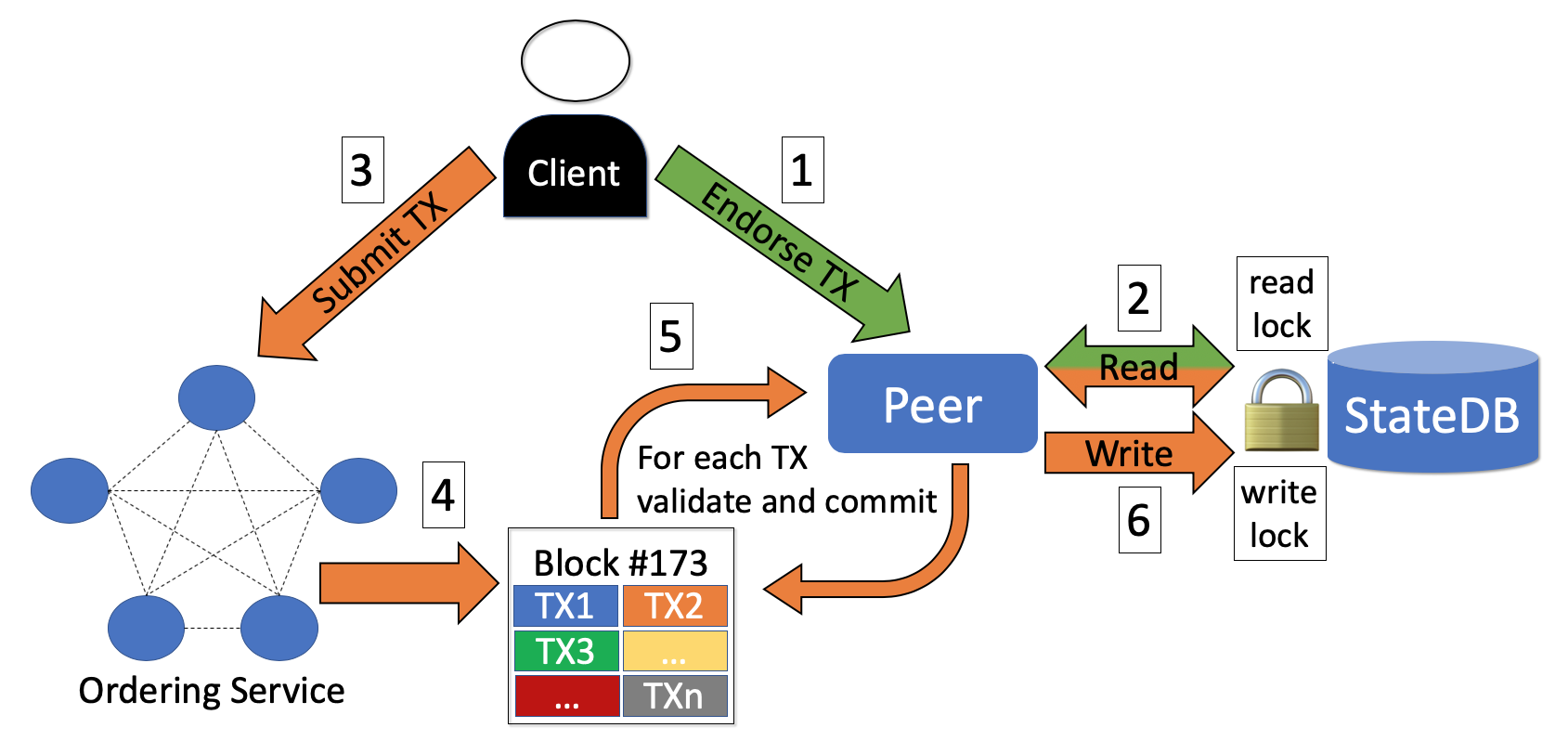}
	\caption{Transaction flow and lock acquisition phases: the client sends transactions to the peer to be endorsed (1), the peer simulates the transactions on its state database while acquiring a read lock (2). Concurrently, the client sends transactions to be ordered into blocks (3) which are then delivered to the peer (4). The peer validates and commits each transaction (5) by updating the state database under a write lock (6).}
	\label{fig:flow}
\end{figure*}

%=== problem definition
The execute-order-validate architecture, as opposed to order-execute, also implies parallelization of transaction execution and validation (including commit).
These phases are performed concurrently by each peer on the blockchain state, implemented by a local key-value store (\emph{state database}).
However, with this parallelization, Hyperledger Fabric also needs to guarantee proper transaction safety, as defined by the ACID properties (atomicity, consistency, isolation, durability)~\cite{acid}.
Currently in Hyperledger Fabric, transaction isolation is attained by locking the entire state database during simulation of transactions. During the endorsement phase, the transaction simulator, which maintains interaction with the state database on the peer, keeps a read lock to allow consistent read of the state keys. During state database update, i.e. during transaction validation and commit phase, the peer keeps a write lock on its state database. This is a shared lock, and contention on that lock significantly impacts performance as indicated by previous work~\cite{thakkar2018performance}.
The transaction flow and lock acquisition phases are illustrated in Figure~\ref{fig:flow}.

This work proposes an alternative approach for achieving transaction isolation while removing the need of the shared lock, thus mitigating the contention during transaction simulation and database update phases. In fact we suggest to utilize enhanced read-write conflict resolution mechanisms to provide a version based snapshot transaction isolation. To this end, we make use of the inherent versioning mechanism of keys in the peer's local state database, used today to perform a read-write conflict detection during the validation phase, and employ it in a novel transaction isolation algorithm.
We show that our method outperforms the shared lock mechanism by a factor of $\sim8.1$. Moreover, we show that our method is comparable to the optimal case where no isolation mechanism is applied, by comparing it to the current implementation with the lock simply removed.

The  rest  of  the  paper  is  organized  as  follows: Section~\ref{sec:background} provides a more detailed background on Hyperledger Fabric and its inner working. The main contribution of our work is presented in Section~\ref{sec:solution} where the proposed lock-free isolation mechanism is described. Section~\ref{sec:evaluation} provides an evaluation of our method. Section~\ref{sec:related} discusses some related work and Section~\ref{sec:conclusion} concludes the paper.

\section{Background} \label{sec:background}
%Background
Prior to Hyperledger Fabric, all blockchain platforms, permissioned (e.g., Tendermint~\cite{Tendermint}, Chain~\cite{Chain}, and Quorum~\cite{Quorum}) or permissionless (e.g., Ethereum~\cite{Ethereum}), followed the order-execute pattern. That is, network participants use a consensus protocol to order transactions, and only once the order is decided, all transactions are executed sequentially, thus essentially implementing active state machine replication~\cite{SMR, replication}. The order-execute approach poses a set of limitations. The fact that transactions have to be executed sequentially effectively leads to throughput degradation, becoming a bottleneck. In addition, an important issue to consider is the possible non-deterministic outcome of transactions. The active state machine replication technique implies that transaction execution results have to be deterministic in order to prevent state ``forks'' which may lead to double spending~\cite{doubleSpending}. Most of the current blockchain platforms implement domain specific languages (e.g., Ethereum Solidity~\cite{Solidity}) to overcome the problem of non-determinism.

Hyperledger Fabric provides a modular architecture and introduces a novel execute-order-validate approach to address the limitations of the order-execute approach mentioned above.
In the execute-order-validate architecture transaction flow is separated into three steps: (1) \emph{executing} a transaction and checking its correctness, thereby \emph{endorsing} it;
(2) \emph{ordering} through a consensus protocol;
and finally (3) transaction \emph{validation} and commit.
Next we explain different aspects of the Hyperledger Fabric architecture.

\subsection{Node types}
The Hyperledger Fabric blockchain network is formed by nodes which could be classified into three categories based on their roles:
\begin{enumerate}
	\item \textbf{Clients} are network nodes running the application code, which coordinates transaction execution. Client application code typically uses the Hyperledger Fabric SDK in order to communicate with the platform.
	\item \textbf{Peers} are platform nodes which maintain a record of transactions using an append-only ledger, and are responsible for the execution of the chaincode and its life-cycle. These nodes also maintain a ``state'' in the form of a versioned key-value store. In order to allow load balancing, not all peers are responsible for execution of the chaincode, but only a subset of peers, the endorsing peers.
	\item \textbf{Ordering nodes} are platform nodes which form a cluster that exposes an abstraction of atomic broadcast in order to establish total order between all transactions. Ordering nodes are completely oblivious to the application state and don't take any part in transaction validation or execution.
\end{enumerate}

\subsection{Distributed application}
A distributed application in Hyperledger Fabric is comprised of two main parts:
\begin{enumerate}
    \item \textbf{Chaincode} is the business logic implemented in a general purpose programming language (Java, Go, Javascript) and invoked during the execution phase. The chaincode is a synonym for the well known concept of smart contracts and is a core element of Hyperledger Fabric.
    \item \textbf{Endorsement policies} are rules which specify what is the correct set of peers responsible for the execution and approval of a given chaincode invocation. Such peers, called \emph{endorsing peers}, govern the validity of the chaincode execution results by providing a signature over these results.
\end{enumerate}

\subsection{Transaction execution flow}\label{sec:flow}
The following summarizes the execution flow of a transaction submitted by a client into Hyperledger Fabric (see Figure~\ref{fig:flow}):

\begin{enumerate}
    \item The client uses an SDK to form a \emph{transaction proposal}, which includes the chaincode name, the function to invoke, and the input parameters to the chaincode function that are about to be executed. Next, the client sends the transaction proposal to the endorsing peers.
    \item Endorsing peers simulate the transaction based on the parameters received from the client. They invoke the chaincode, record state updates, and produce output in the form of a versioned read-write set. The state does not change at this stage. Next, each endorsing peer signs the read-write set and returns the result back to the client.
    \item The client collects responses from all endorsing peers, validates that results are consistent, i.e. all endorsing peers have signed the same payload. It then concatenates all  signatures and identities of the endorsing peers along with the read-write sets, and assembles them into a transaction which is submitted to the ordering service. 
    \item The ordering service batches the transactions into blocks, which induces total order among transactions within and outside the same blocks. The blocks are then delivered to the peers.
    \item Upon receiving a new block, each peer iterates over the transactions in it and validates: a) the endorsement policy, i.e. whether the set of endorsing peers signatures satisfies the endorsement policy of the corresponding chaincode; b) performs multi-version concurrency control (MVCC)~\cite{mvcc} check against the state.
    \item Once the transaction validation has finished, the peer appends the block to the ledger and updates its state based on the valid transactions. After the block is committed the peer emits events to notify clients connected to it.
\end{enumerate}

\subsection{Locally stored state}
Each peer uses a local key-value store, implemented with either LevelDB (in Go)~\cite{LevelDB} or Apache CouchDB (single, non-clustered, instance)~\cite{CouchDB}, to maintain the latest state, in addition to a locally stored ledger.
For each unique key the peer stores in the database a key-value pair of the form $\big(key, \langle val, ver\rangle\big)$, containing the key's most recently stored value \emph{val} and its latest version \emph{ver}.
The version is a monotonically increasing number which represents the last transaction that updated the key.
It consists of the block sequence number and the sequence number of the transaction within the block $\big(\langle BlockNum , TXNum\rangle\big)$.

Additionally, Hyperledger Fabric peers store a key in their state database termed as a \emph{savepoint}.
The key's value is of the form $\big(\langle BlockNum , TXNum\rangle\big)$, same as the form of versions of other keys in the state database.
This value is updated at the last part of each validation phase, after the entire block is committed to the state database.
The number of the last transaction in the block and the block number itself are written as the new value of the savepoint.
And so, the savepoint's value essentially represents the highest key's version in the state database at this time, the last modifying transaction.

As mentioned in Section~\ref{sec:flow}, during the execution phase peers simulate the transaction and produce a \emph{writeset} - modified keys with their new values, and a \emph{readset} - all keys read during simulation with their versions.
In the validation phase peers perform a MVCC check on readsets of all transactions in the block sequentially.
For each transaction it compares the versions of the keys in the readset to those in the current state, and ensures they are still the same.
If the versions do not match, the transaction is marked as invalid and its effects are disregarded.
Finally, all state updates of valid transactions are applied by writing all key-value pairs in writesets to the database.

The transaction execution step and the block validation and commit step are performed in parallel by a peer, as enabled by the execute-order-validate architecture.
However,  in the current implementation of Hyperledger Fabric, to ensure transaction isolation, in the validation phase, an exclusive write lock is acquired on the whole database, and during execution phase a shared read lock is acquired. The lock acquisition phases are demonstrated in Figure~\ref{fig:flow}. 

\section{Proposed solution} \label{sec:solution}
%solution
In our proposed solution we utilize the native versioning of the keys in the state database, used primarily to perform the MVCC check during the validation phase, together with the already existing savepoint mechanism, to create a consistent view of the state for the simulations. 
The main insight is that key-value versions are not incremented independently, but rather in a database-wide monotonically increasing manner, representing transaction commit order.
Moreover, the savepoint represents a boundary between transactions. 
Our solution uses the savepoint during simulation in order to detect whether the simulated transaction is violating the transaction boundary of a concurrently committing transaction.
If we detect there might be an isolation violation, we abort the simulation. 
This enables us to remove the shared lock, eliminating contention, and still assure the isolation of the transactions.

Next we detail our proposed isolation solution.
Section~\ref{sec:basic} focuses on our basic algorithm, while in Section~\ref{sec:MVCC} we discuss the relation between simulation aborts and MVCC check failures.
In Section~\ref{sec:deleted} we specify how we deal with deleted keys.
Finally, Section~\ref{sec:rich} describes the difficulty presented by rich queries.

\subsection{The basic algorithm}
\label{sec:basic}

The basic solution (Algorithm~\ref{alg:sim}) records the current savepoint at the beginning of a simulation (line~\ref{line:record}).
A key is read during the chaincode execution by invoking the GET procedure (line~\ref{line:getProcedure}).
After obtaining the state of the key from the database (line~\ref{line:getDB}), if the key appears in the database, we check that its version is equal to or less than the recorded savepoint.
If the version is greater than the recorded savepoint, then we abort this transaction (line~\ref{line:verCheck}).
Otherwise, we return the key's value (line~\ref{line:returnVal}). Any other aspects of the transaction endorsement phase remain intact.

\begin{algorithm}[t]
\small
\caption{Simulation}
\label{alg:sim}
\begin{algorithmic}[1]
\State \texttt{savepoint} $\gets \bot$

\item[]

\Procedure{begin}{}
\State \texttt{savepoint} $\gets$ DB.GetSavePoint() \label{line:record}
\EndProcedure

\item[]

\Procedure{get}{\texttt{key}} \label{line:getProcedure}
\State $\langle$\texttt{val},\texttt{ver}$\rangle$ $\gets$ DB.GetState(\texttt{key}) \label{line:getDB}
\If{$\langle$\texttt{val},\texttt{ver}$\rangle \neq \bot$} \label{line:verNil}
\If{$\texttt{ver} > \texttt{savepoint}$} \label{line:verCheck}
	\Return \textsc{error} \Comment{abort}
\EndIf
\EndIf
\State \Return \texttt{val} \label{line:returnVal}
\EndProcedure
\end{algorithmic}
\end{algorithm}

As an example, consider the following scenario:
Let tx1 and tx2 be two transactions and assume that tx1 performs a read of key A after which it performs a read of key B, while tx2 performs a put to key A and a put to key B.
If we commit tx2 during the simulation of tx1, between the read of key A and the read of key B, then without locking the database, isolation is breached and correctness is impaired, therefore, we need to abort the transaction execution.
Since the read of key A returned before the commit of tx2, with an approved version, and the read of key B returned after, with a version definitely greater than the recorded savepoint, in our proposed solution tx1 is aborted.
This transaction abort example is illustrated in Example~\ref{abort}.

\begin{table}[b]
	\centering
	\begin{tabular}{c|c|c|c|c|c}
		\hline
		\bfseries Time & \bfseries Tx & \bfseries Op & \bfseries Key & \bfseries Val & \bfseries Ver\\
		\hline
		1 & tx1 & read & savepoint & $\langle100,275\rangle$ & $\langle100,275\rangle$\\
		2 & tx1 & read & A & 20 & $\langle100,250\rangle$\\
		3 & tx2 & put & A & 21 & $\langle101,345\rangle$\\
		4 & tx2 & put & B & 47 & $\langle101,345\rangle$\\
		5 & tx1 & read & B & 47 & $\langle101,345\rangle$\\
		6 & tx1 & abort & & & \\
		\hline
	\end{tabular}
	\bigskip
	\caption{\textnormal{Simulation abort}}
	\label{abort}
\end{table}

This algorithm guarantees that the simulation gets a consistent view of the state, previously assured by locking.
In this view, every key read was updated before the savepoint was recorded.
If a simulation encounters a key that has a greater version than the recorded savepoint, then we say that it breaks the consistency of the view.
Such a read may puncture the isolation of the simulation, since the update could have happened during the simulation, and so we perform an abort.

Furthermore, the algorithm assures the atomicity of transactions without the use of locking.
Only after the commit of the entire block, the savepoint is updated with the latest committed transaction number and block number.
The version check in Algorithm~\ref{alg:sim} causes an abort in cases where the key's version is higher than the recorded savepoint at the beginning of the execution, which implies that the key may be from a not fully committed transaction (otherwise, the recorded savepoint would've been at least as big as the key's version, and the transaction is assured to be fully committed).
Therefore, the atomicity of transactions is preserved.

\subsection{Aborts and MVCC check failures}
\label{sec:MVCC}
One point to notice is the relation between aborts and MVCC check failures.
Assume there is an execution of a transaction that is later validated and committed.
If the execution aborts then this means there was a read preceded by an update to a key by a concurrent commit.
If the commit process starts after the execution begins, and if a simple reader/writer mutual exclusion lock is used, then this commit is scheduled after the execution.
In this case, there is no abort (since lock is used), and after the transaction executes it reaches the validate and commit phase.
Then the transaction goes through a MVCC check where it fails since there was an update (by the earlier commit) to the key that the execution read.
Therefore, in this case, an abort is essentially an early MVCC check failure detection. Instead of having the transaction go through all of the phases just to fail the MVCC check, it aborts at execution time, which improves overall system throughput.

\subsection{Deleted keys}
\label{sec:deleted}
A key deletion removes the key from the database completely.
Consider the case of a key removed from the database after a simulation starts and before a read of the same key.
If we examine the basic algorithm (Algorithm~\ref{alg:sim}), in this case the version check is not performed (line~\ref{line:verNil}), and the simulation does not abort.
However, the isolation of the transaction might be broken and so this simulation should be aborted.

Therefore, instead of removing keys entirely from the database at a deletion, we keep deleted keys in the database.
Furthermore, we update the deleted key's version to indicate the transaction that performed the deletion (same as any other key update).
To differentiate between deleted keys and non deleted keys, we add a flag to each key which indicates whether this key is deleted or not.
This flag can be implemented using the MSB of the transaction number in the version (since there is a limited number of transactions inside a block), and so reading this flag does not require more database requests.

We cannot leave the deleted keys in the database, since actions like range queries might become slower.
Therefore, we create a \emph{garbage collector} (GC) to remove deleted keys from the database.
We maintain a list of deleted keys, that is updated at commit time and the GC (called every few commits) examines the list to decide which keys to remove entirely from the database.

We don't remove deleted keys that can potentially be read by an ongoing simulation, that started before the deletion of the key, since this may break the isolation.
Therefore, we keep track of the recorded savepoints of all ongoing simulations.
The GC is allowed to remove a key only if the key's version is less than all ongoing simulations savepoints, meaning the key was deleted prior to the start of all ongoing simulations and therefore removing it cannot harm the isolation.

\subsection{Rich queries}
\label{sec:rich}

CouchDB, unlike LevelDB, supports rich queries by enabling document search using a declarative JSON querying syntax. Isolation of simulations with rich queries is not possible when considering a solution that involves tracking the queried keys, as done in our solution.

\begin{lstlisting}[float=htb, basicstyle=\small, caption={Cars txs}, captionpos=b, label={example:carsTXs}, frame=single]
TX1:
query(cars with owner=Alice)
query(cars with owner=Bob)

TX2:
put(car1, owner=Bob)

TX3:
put(car1, owner=Alice)
\end{lstlisting}

\begin{figure}[htb]
\centering
\includegraphics[width=\linewidth]{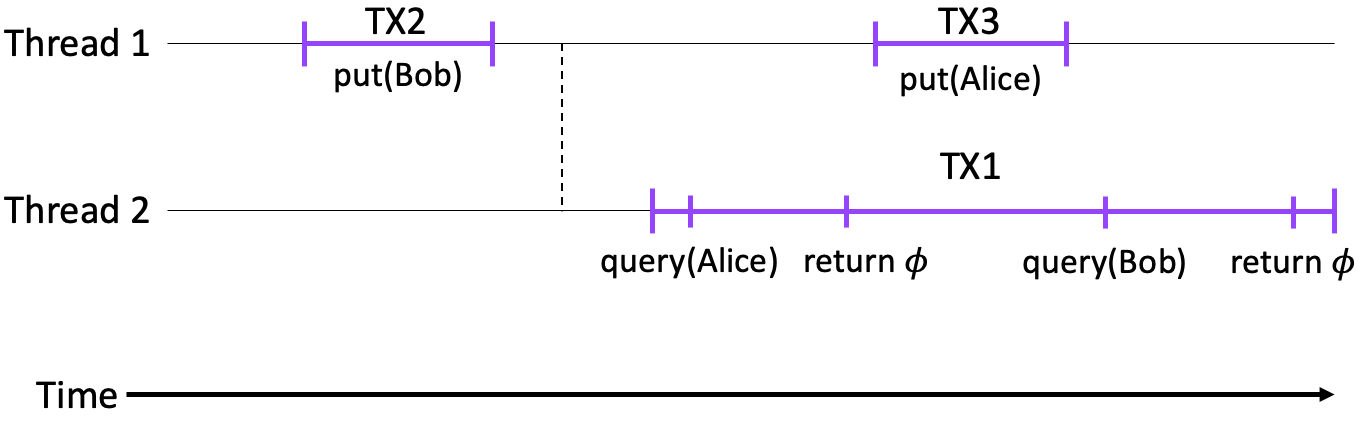}
\caption{Cars scheduling}
\label{fig:carsScheduling}
\end{figure} 

Consider a car rental system, where each key represents a car and its value contains information such as the model, year, color, owner, etc.
Now we compose  3 different transactions, one with rich queries, looking up cars with a specific owner, and two others that update a car's owner, illustrated in Example~\ref{example:carsTXs}.
Next we examine a scenario where queries are executed while the updating transactions are committed.
Figure~\ref{fig:carsScheduling} presents such a scenario, it shows how a car (car1) can ``disappear'' and execution does not abort.
This transactions scheduling is not serializable~\cite{TransactionalSerializability,DataBaseSerializability}, since there is no equivalent serial execution.
For each query the database returns with an empty set, and no keys are entered into the read set.
The version check passes, because it is applied only on the returned keys from the query (an empty set), and consequently the transaction isolation is not attained.
The fact that the database returned an empty set for the query and not all the cars in the database is the reason why our proposed solution does not work.

\section{Evaluation} \label{sec:evaluation}
%evaluation
We implement the proposed solution on top of Hyperledger Fabric v1.2.
All experiments were run on a virtual machine with an Intel(R) Xeon(R) CPU E5-2650,
16 GB of memory, and a SSD with measured write throughput of $\sim$~320~MBps.

To run experiments we use a customly built tool, that embeds a Hyperledger Fabric client implementation in Golang \cite{goSDK}.
This tool sends transactions to a peer to be executed and endorsed, then stores the assembled transactions in an in-memory pool.
Concurrently, the tool sends transactions stored in the in-memory pool to the orderer.
The time it takes for transactions to be executed and endorsed by the peer is measured and reported by the tool.
The number of transactions and concurrency level is parameterized and can be controlled.
We write a chaincode to run experiments using this tool.
Each experiment consists of 10 iterations and we report the average result.

\subsection{Performance}

We compare our solution to the current (\emph{lock-based}) implementation in Hyperledger Fabric and an \emph{optimal} implementation where the lock is simply commented out.
The optimal implementation can be used only with workloads where transaction isolation is assured (such as read-only and write-only workloads).

\begin{figure}[htb]
\centering
\includegraphics[width=\linewidth]{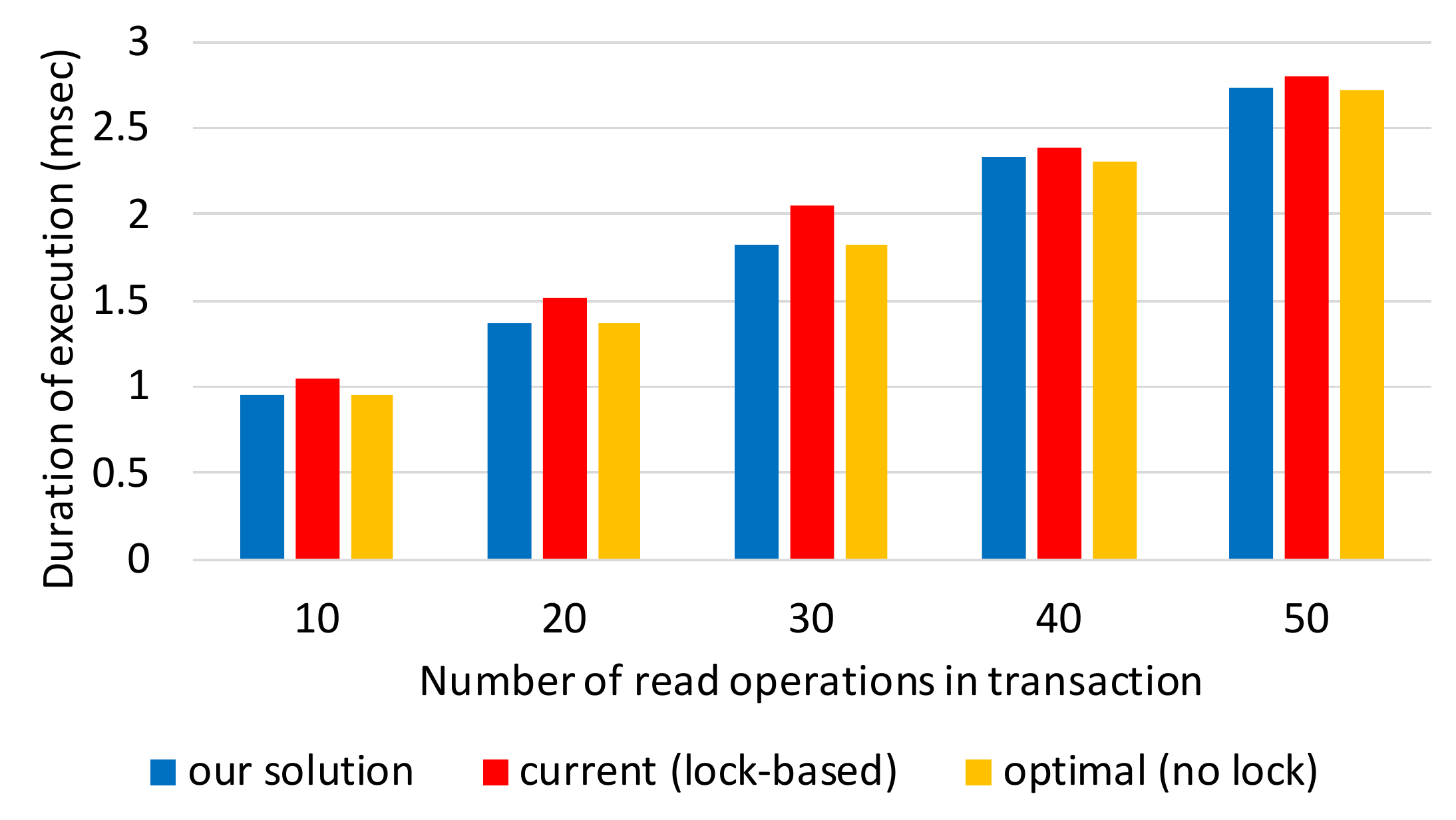}
\caption{Read-only workload}
\label{fig:readonly}
\end{figure}

The first experiment is a read-only workload with LevelDB, where simulations of read-only transactions are done concurrently with commits of read-only transactions.
Figure~\ref{fig:readonly} shows the duration of an execution (in milliseconds) in the read-only workload.
When we increase the number of read operations in the transactions  we can see that the time it takes to endorse grows for all implementations.
In this case the commit time is low since there is no updates to commit (as this is a read-only workload), and so the lock is not held for a long time.
Therefore, the lock causes just a minor overhead (0.1 ms), as presented by the difference between the lock-based and the optimal execution duration.
Moreover, our solution execution duration is very close to the optimal execution duration and slightly less than the lock-based duration.

\begin{figure}[htb]
\centering
\includegraphics[width=\linewidth]{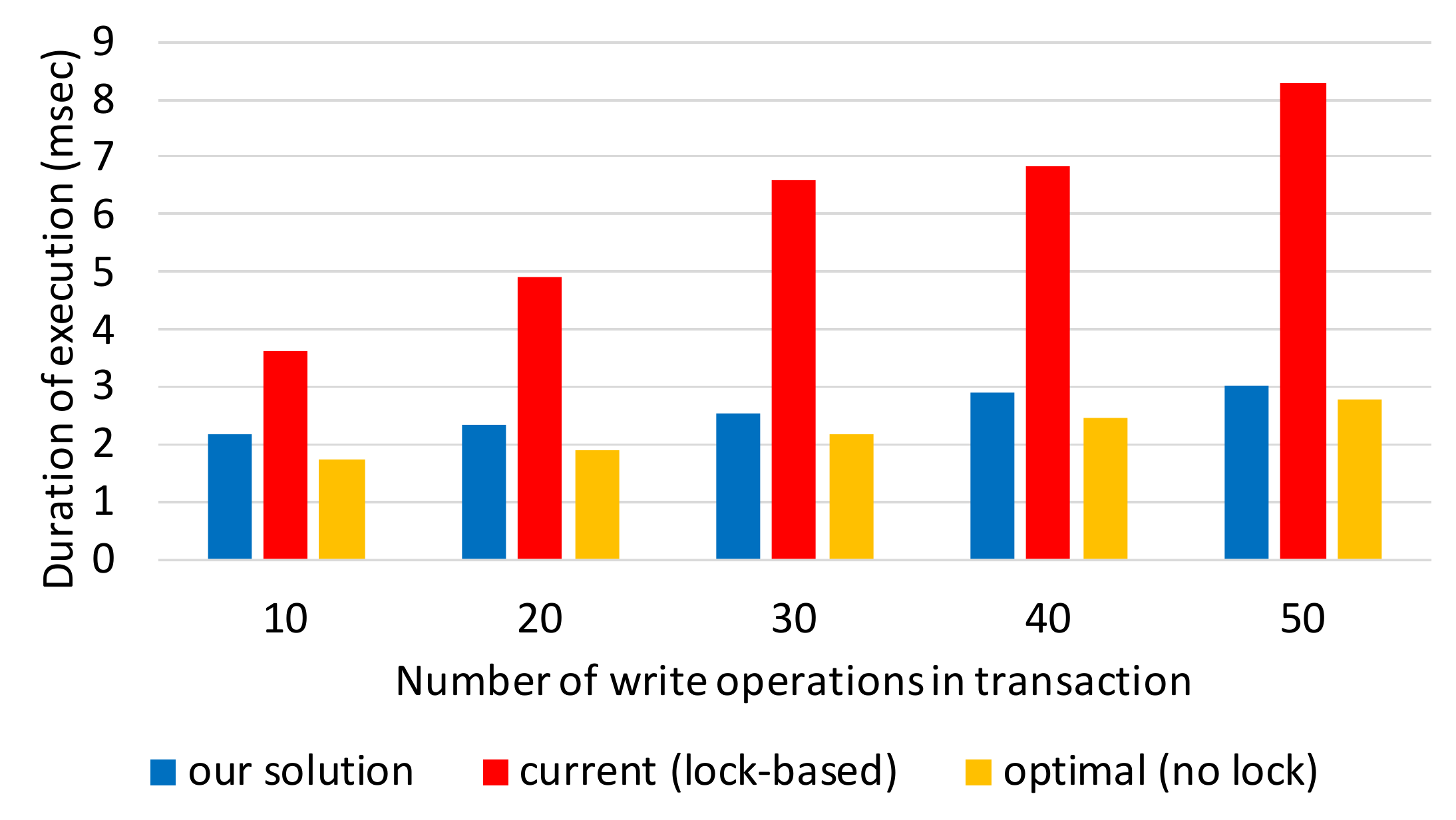}
\caption{Write-only workload}
\label{fig:writeonly}
\end{figure}

\begin{figure}[htb]
\centering
\includegraphics[width=\linewidth]{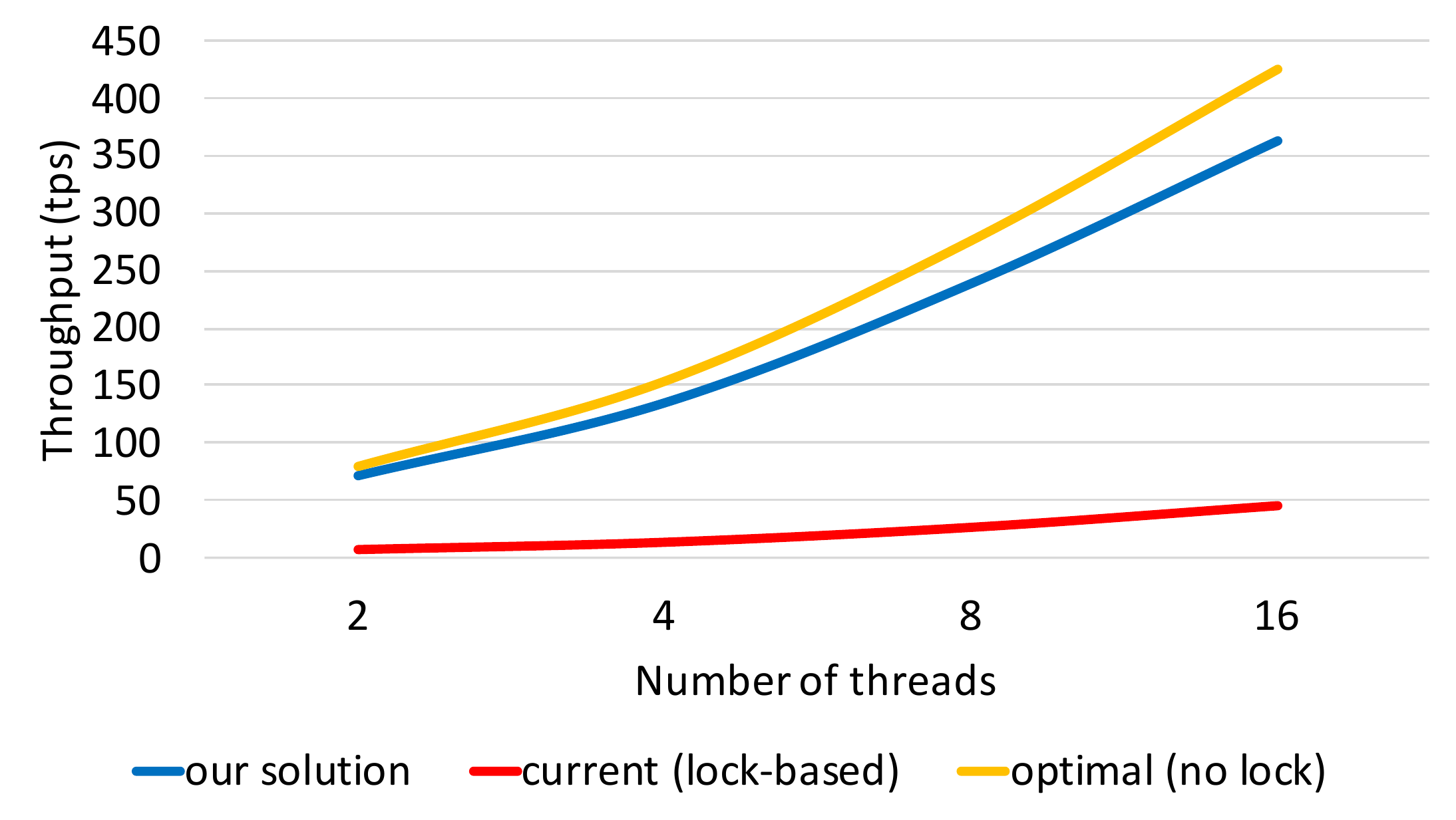}
\caption{Throughput scalability with the number of threads}
\label{fig:throughput}
\end{figure}

The second experiment is a write-only workload with CouchDB.
In this case the commit takes a long time since there are multiple updates to CouchDB, therefore, in the current implementation, the lock is held for a long time which causes a big overhead.
Figure~\ref{fig:writeonly} shows the difference in execution time for the write-only workload.
In the worst case (50~writes), lock-based takes 8.3~ms while optimal takes only 2.8~ms (speedup of~3x), and our solution takes just 3~ms (2.8x~speedup from lock).
With 10~writes, lock-based takes 3.6~ms, optimal takes 1.7~ms (2.1x~speedup), and our solution takes~2.2 (1.6x~speedup from lock-based).
Here too our solution execution duration is very close to the optimal execution duration and far less than the lock-based duration.

In the next experiment we measure the throughput scalability with the number of threads.
We run a write-only workload with CouchDB where a 1000 transactions (each transaction consists of 10 write operations) are executed while commit occurs concurrently.
Figure~\ref{fig:throughput} depicts the throughput (in transactions per second) in the write-only workload. Our solution outperforms the current lock-based solution by 8.1x and optimal outperforms lock-based by 9.5x (not far from our solution).

\section{Related work} \label{sec:related}
%related
In blockchain networks all non-faulty (honest) nodes must maintain a consistent view of the ledger and the world state, therefore every node has to process transactions in the exact same order. This is achieved through a consensus algorithm, which imposes order between transactions and forms blocks that are hash chained together to eventually construct the blockchain.

Each blockchain transaction operates on the world state, and so it is required to be isolated.
In fact there is a need to prevent dirty, phantom, or non-repeatable reads~\cite{berenson1995critique}. Moreover, transactions should conform to all of the ACID properties~\cite{acid}.
This could be achieved by means of optimistic concurrency control~\cite{optimisticConcurrencyControl}, serializable snapshot isolation~\cite{ssi, ports2012serializable}, or with two-phase locking~\cite{twoPhaseLocking}.

Concurrent access to the database without a proper control mechanism may put data integrity at risk. In many cases, this is handled by locking mechanisms which provide exclusive access to the data that is being mutated~\cite{menasce1980locking}. Alternatively, snapshot isolation or optimistic concurrency control are known techniques that achieve serializability.

The locking approach comes with its own disadvantages, the most prominent being that locking, in its essence, diminishes concurrency and therefore reduces throughput. In addition, locking incurs maintenance overhead during the system's development, in order to ensure that the locking does not lead to deadlocks and preserves integrity. Sometimes, complex deadlock detection is employed to ensure recovery from unforeseen deadlocks.

To mitigate these problems, some systems adopt an optimistic approach \cite{optimisticConcurrencyControl} to locking.
The main idea behind optimistic concurrency control, is to avoid needless locking and by doing that - increase the throughput of the system at a whole. It puts faith in data dependency conflicts between parallel running transactions not to occur, and structuring the transaction's life-cycle into a 3-step process:
\begin{itemize}
	
	\item Read step - Data objects that are accessed by the transaction are tracked during the transaction's execution, as well as the type of access (read, write, delete). A copy of the data object is made upon the first update, and subsequent updates to the object are applied to the copy.
	
	\item Validation step - The goal of this step, is to provide "serial equivalence" to the transaction
	
	\item Write step (in case the transaction updates data) - Atomically swaps the global data objects with the copies updated during the read phase.
	
\end{itemize}

Our solution was inspired by the transactional locking II (TL2) algorithm~\cite{tl2} which is an opportunistic speculative execution algorithm in which there exists a shared global version clock, and each data field has a lock and a version attached to it as well. TL2 use the data field locks to assure the write step is atomic and to allow multiple threads to perform the write step concurrently.
In our solution, there is no need to lock because in our case, the concurrent updates are replaced with a series of database updates, the committing of blocks. In addition, the global version clock is dictated by the savepoint.

Increasing concurrency in blockchains was recently considered also in the context of order-execute architecture~\cite{addingConcurrency}, where execution of smart contracts in parallel can be accomplished by using techniques adopted from software transactional memory. To the best of our knowledge, this work is the first to offer such increased concurrency in the execute-order-validate blockchain architecture. 

\section{Conclusion} \label{sec:conclusion}
%conclusion
This work presented a new lock-free approach for providing transaction isolation in Hyperledger Fabric.
The method utilizes the native versioning of keys in the peer's state database, and makes use of the fact that the versions are database-wide monotonically increasing, representing transaction commit order.
The method then exploits the \emph{savepoint} as a boundary between transactions, and uses it to detect whether the simulated transaction is violating the transaction boundary of a concurrently committing transaction.
When an isolation violation is detected, the transaction simulation is aborted. 
This technique allows us to remove the shared lock, eliminating contention, while still ensuring transaction isolation.

Our experiments have shown that the proposed solution achieves better throughput and lower simulation latency than the current lock-based implementation of Hyperledger Fabric. The performance gain is especially high in write intensive workloads. 
In addition, our results demonstrated that the performance achieved by our technique is comparable to the optimal, which is approximated by a simple disposal of the lock. 

\bibliographystyle{IEEEtran}
\bibliography{references}

\end{document}